# Flat Field Forensics

Timothée Greffe*[1] and Roger Smith[1]

[1] California Institute of Technology, MC11-17, Pasadena CA 91125, USA
*tgreffe@caltech.edu

**Abstract.** We present two subtle charge transport problems revealed by the statistics of flat fields. Mark Downing has presented photon transfer curves showing variance dips of order 25% at signal levels around 80% of blooming. These dips appear when substrate voltage is raised above zero, for -0V to 8V parallel clock swing. We present a modified parallel transfer sequence that eliminates the dip, based on the hypothesis that it is caused by charge spillage from last line to the 2nd last line. We discuss an experiment to test whether the electrode map is incorrectly reported in the data sheet. A more subtle dip in the variance occurs at signals around 6000 e-. This is eliminated by increasing serial clock high by a few volts, suggesting the existence of a small structural trap at the parallel-serial interface. Tails above blooming stars are suppressed using an inverted clocking during readout and a positive clocking during exposure to maintain sharpness of the PTC. We show that integrating under three parallel phases, instead of the two recommended, reduces pixel area variations from 0.39% to 0.28%, while also eliminating striations observed along central columns in pixel area maps. We show that systematic line and column width errors at stitching boundaries (~15 nm) are now an order of magnitude less than the random pixel area variations.

**Keywords:** WaSP, Zwicky Transient Facility, CCD-231, Teledyne-e2v, PTC, Downing Dip, charge smearing, Brighter-Fatter effect.

## 1 WaSP Camera

The Wafer Scale Primary Camera (WaSP) is a wide field CCD imager for the P200 telescope at the Palomar observatory. A 6k by 6k CCD231-C6 made by Teledyne-e2v provides uninterrupted coverage of the majority of the 24 arcmin diameter field of view at prime focus (unlike the 6 CCD mosaic that it replaces), while two 2kx2k STA CCDs operated in frame transfer mode are located along one chord to provide Guiding and Focus sensing.

CCD231-C6 data sheet promises 5e- read noise at 1MHz (single sided), QE above 90% over much of the wavelength range and Well Capacity ~350ke-. Only the 2 upper amplifiers are used in the work described here. The sensor is driven by an Archon controller from Semiconductor Technology Associates. Another 16 of the CCD231-C6 detectors are also used on the Zwicky Transient Facility (ZTF) Camera.

## 2 Downing Dip

### 2.1 Flatness of the Photon Transfer Curve

The Photon Transfer Curve (PTC) is computed using only two successive frames with a signal staircase produced by constant illumination during fast readout of a burst of rows with a selectable integrating time between each burst. We call this “movie mode” since it was originally developed to capture a rapid series of images of stars to allow measurement of image motion. The spatial variance of the difference of each signal band divided by √2 is plotted versus the corresponding means. The full of signal range may be covered in just two frame times, if illumination and integrating time are well selected.

The more conventional method is to record two exposures at each illumination level. One pair of frames only yields a single data point. Although this method requires much more storage capability and readout time, it is valuable to verify that the PTC derived from “movie mode” is valid and may be required when cross-correlation, auto-correlation or Fourier transforms are required to investigate anomalous dips in the PTC, especially when behavior depends on position within the image.

Below is a plot of the PTC of the CCD231 when using the recommended 0V to 10V voltages and clocking from e2v. The large number of points on this curve are obtained by using the movie mode with many signal steps.

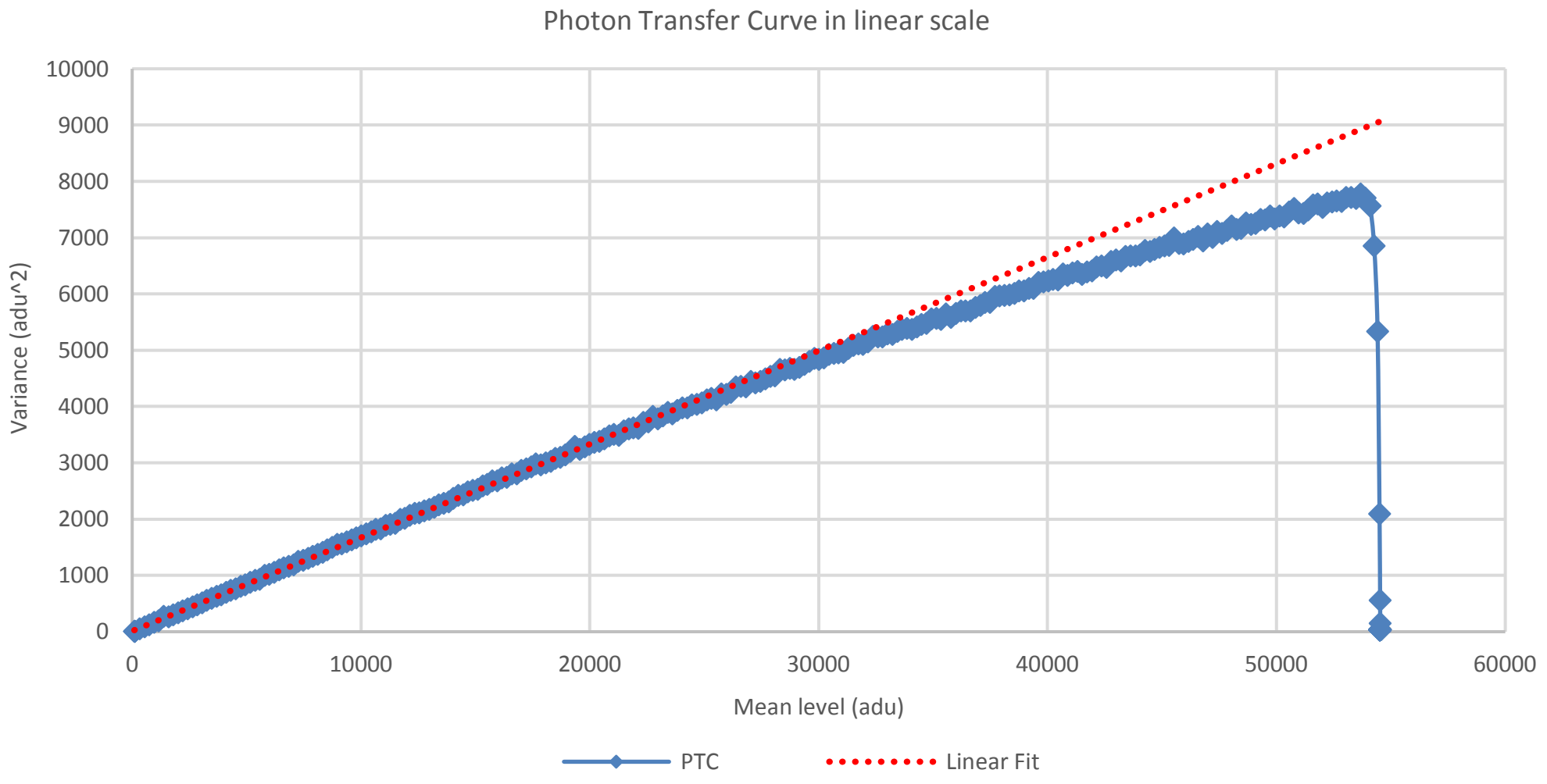

**Fig. 1.** Photon Transfer Curve of the e2v 231 using the recommended voltage.

“Conversion gain” defined as the inverse of slope at low signals, is about 6e-/adu. The well capacity is the maximum value of the mean (~55 kADU) times the conversion gain, 330ke-, which is acceptable.

The reduced slope in the PTC at high signals is not consistent with a change in gain, since signal is linear with exposure time for constant flux to better than 1%.

Astier et al. [1] account for this, calling it the “Brighter-Effect”. Even when mean flux is uniform, Poisson noise causes more charge to accumulate in some pixels than in their neighbors. As this occurs a transverse electric field is created by the uneven distribution of stored charge causing photogenerated electrons near the boundaries to be pushed towards the pixels with less charge. The total amount of charge remains constant but correlation between pixels is created and the spatial variance decreases.

James Janesick [2] describes how the charge cloud under the storage phase can reach the surface before blooming if the storage potential is sufficiently higher than the barrier potential. The onset of trapping at the oxide interface causes charge to be removed from one pixel and released into later pixels. The charge mixing in the “surface full well” condition can also reduce the variance and thus the slope of the PTC. Thus there is some uncertainty regarding the PTC curvature which can be cause by either surface full well or the brighter-fatter effect, or both.

The ultimate precipitous drop in variance occurs either when blooming sets in or the signal chain or ADC saturates. We have setup the gain and offsets so that blooming occurs before ADC saturation

## 2.2 Downing Dip.

PTC for the full range of high and low parallel clock voltage were measured to search for the Optimum Full Well following the method of James R. Janesick [2, page 277], who predicts that useable well capacity is maximized when blooming occurs just prior to surface full well. Not only was the differential swing explored, but also the common mode range. Voltages on the serial register were adjusted to assure efficient transfer from parallel to serial register.

While conducting this search, a class of PTC with unexplained dips appeared at low common mode voltage and relatively low differential voltage. A similar dip was first described by Mark Downing in [3] but left unexplained.

Even though these dips appear only when operating the sensor outside the recommended voltage range, further study of this phenomenon seemed worthwhile since it might expose other important charge transport phenomena. This proved to be the case.

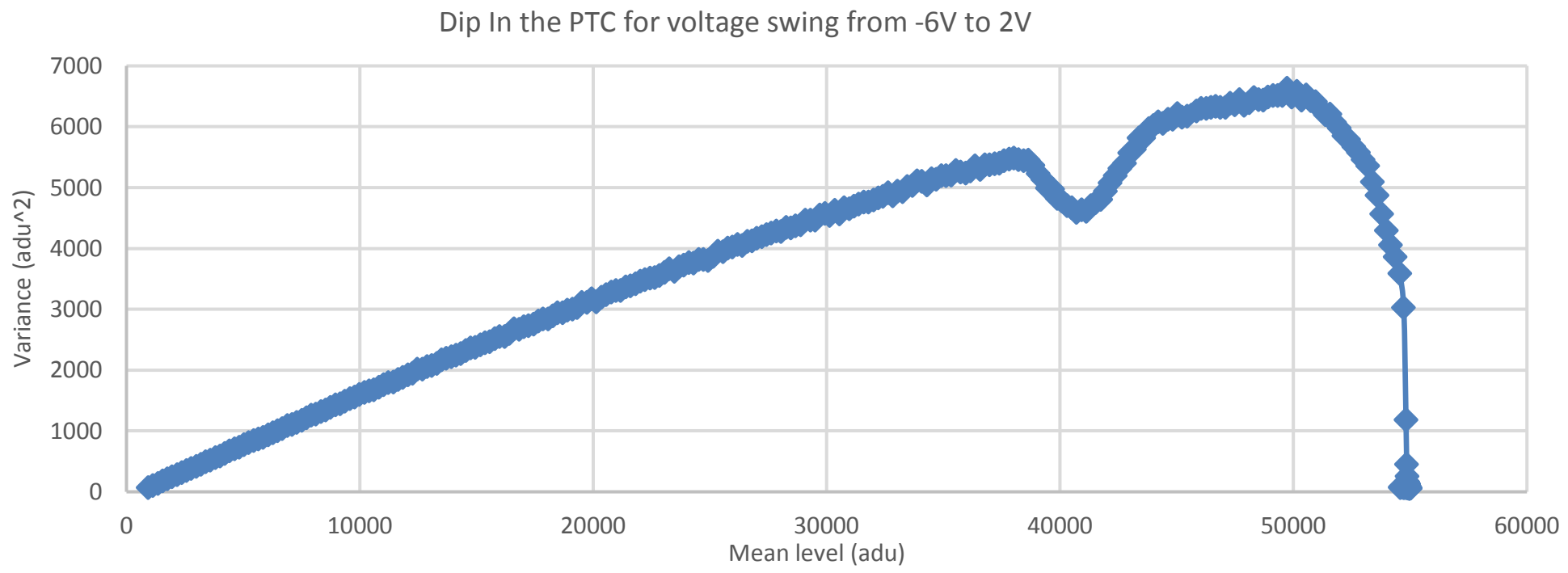

**Fig. 2.** Extract of a Photon Transfer Curve of the e2v 231 showing the so-called *Downing Dip*. The parallel voltage swing from -6V to 2V should deliver a straight PTC over the full range until blooming. What causes this anomaly?

## 2.3 Image shift hypothesis

The noise statistics should follow a Poisson distribution. It is not clear how noise would decrease for a given level, and it is even harder to explain why variance would recover for higher intensities, when no level-sensitive threshold effect is known during the charge transfer.

Some (as yet unidentified) phenomenon causing a drift of the whole image area could explain such behavior. We posit a toy model in which at some signal level, part of the charge in a pixel is delayed by a line the partial summation of charges with that in the following line causes noise correlation and a drop in variance. Below the threshold level the charge is unaffected. Well above the threshold all the charge in a given pixel image has been delayed by one line. The image has been shifted by a line but is essentially intact so the variance is unaltered.

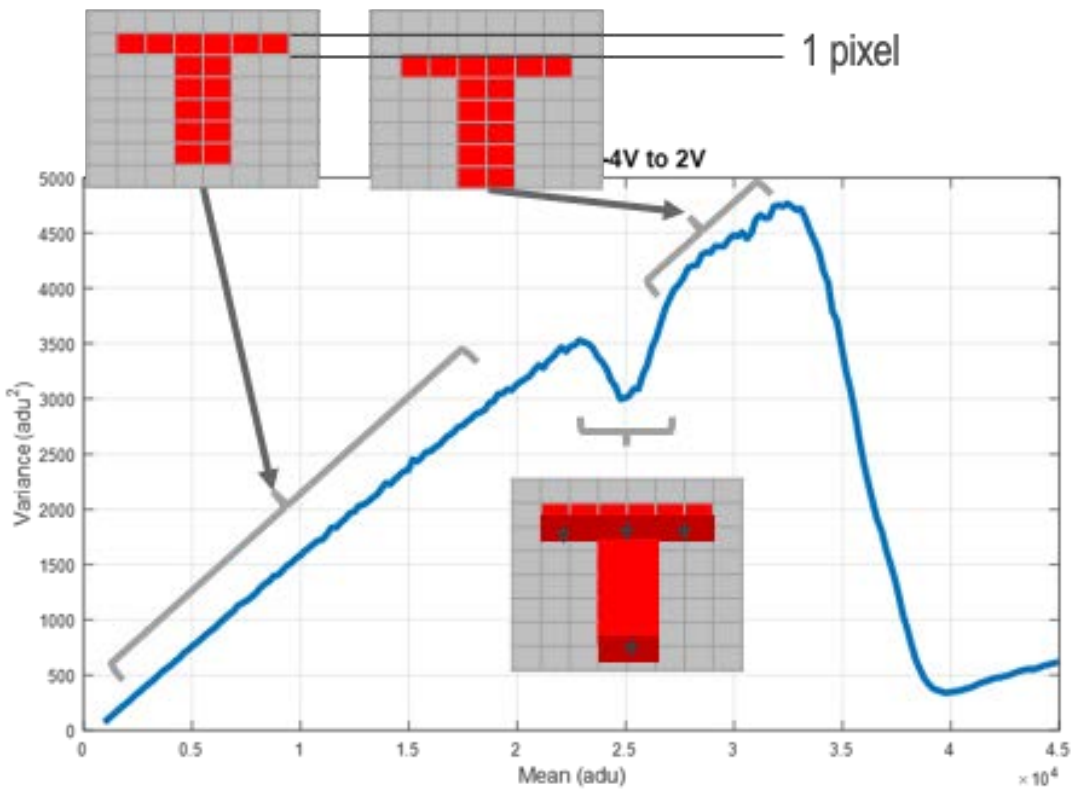

**Fig. 3.**image shifting theory proposed to explain the dip. Parallel clock voltage between -4V and 2V.

*Considering a frame C made of the complementing sum of frames A and B.*

$C = (1-\eta)A + \eta B$ *with* $\eta \in [0;1]$

*We can show that the resulting noise of C is:*

$$Var((1-\eta)A + \eta B) = (1-\eta)^2\sigma_A^2 + \eta^2\sigma_B^2 + \eta(1-\eta)Co\,var(A,B)$$

*Where the last term is null in 0 and 1 and represents the noise correlation between pixels of a same frame.*

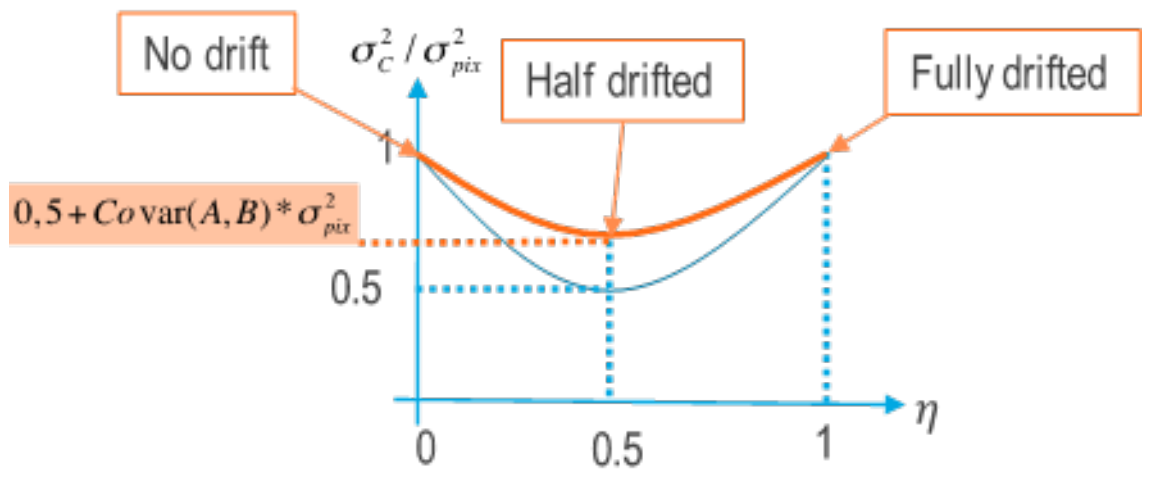

**Fig. 4.**image shifting theory proposed to explain the dip. Analytical justification.

Is this actually happening? Image shifts on sub pixel scales can be accurately measured by cross-correlation even with flat illumination since there is typically a fixed pattern in the response with 0.5% rms variation on spatial scales of a few pixels, which is typically due to pixel area variability [4] as well as localized surface features such as dust spots[5]. In **Fig. 5** the reference frame is taken on the first part of the PTC. The center of gravity is calculated for the Gaussian cross correlation profile and then plotted with the PTC. This method relies on the Fixed Pattern Noise of the sensor to show image registration.

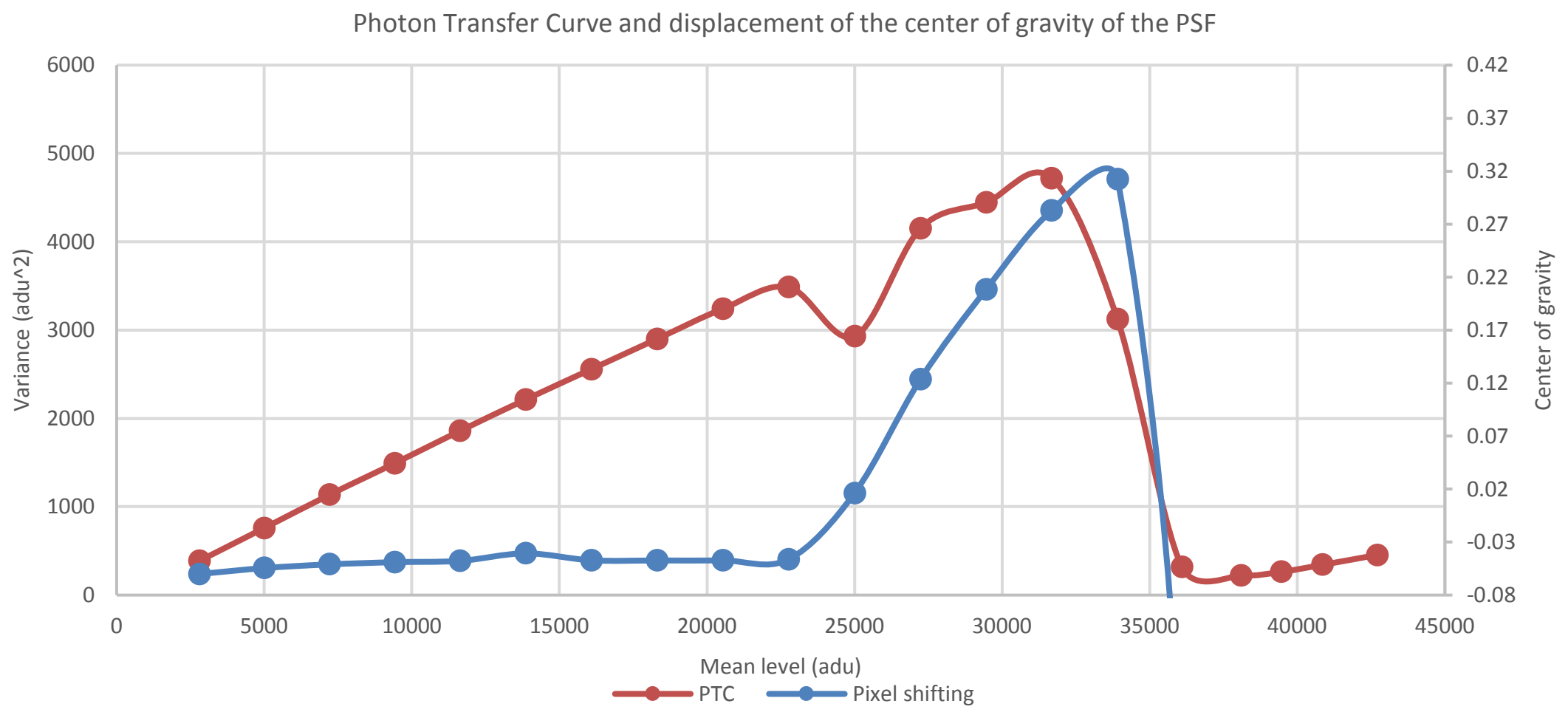

**Fig. 5.** Dip in the Photon Transfer Curve and shifting of the pixels.

This graph demonstrates that there is indeed a shift of the image area but not until the intensity at which the Downing dip occurs .

The shift in the centroid of the cross-correlation, is significantly less than one pixel. It is unclear why.

It is also possible to simply look at the value of the first *parallel* overscan rows to find the same result. However, the cross-correlation tool is more powerful because it shows that the shifting occurs everywhere on the image area by selecting different sub-windows for the processing.

This result is not obvious and leads to the thought that the shifting mechanism itself is occurring at the border between the image area and the serial register, and not inside the image area either progressively at each line transfer or abruptly at some edge within the image area.

### 2.4 Proposed mechanism

Since the pixel shifting occurs at the edge of the image area, we focused on the interface between last two lines and the serial register. **Fig. 6** shows how charge should be transferred from one pixel to the next one and then finally to the serial register.

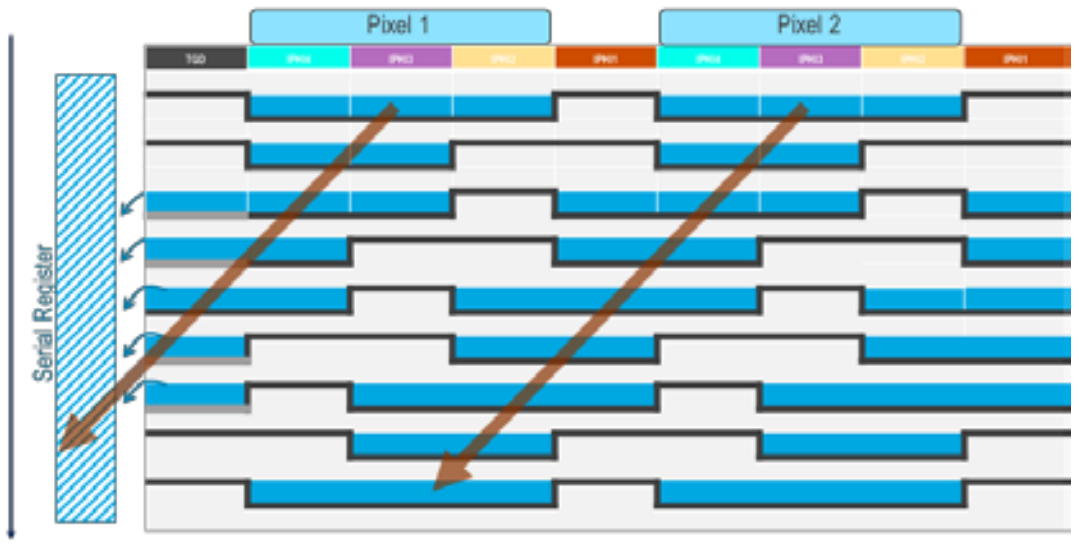

**Fig. 6.** Charge transfer on the Teledyne-e2v CCD-231-C6 using the recommended voltage and clocking. Charges are displayed in blue and the pixels are moved from a row to another during a cycle.

**Fig. 6** does not explain a pixel shifting at some signal threshold. Could the last node before the transfer gate be missing or even be shorted to its neighbor? A new clocking diagram is sketched below to explain how the charges would then flow from the image area to the serial register.

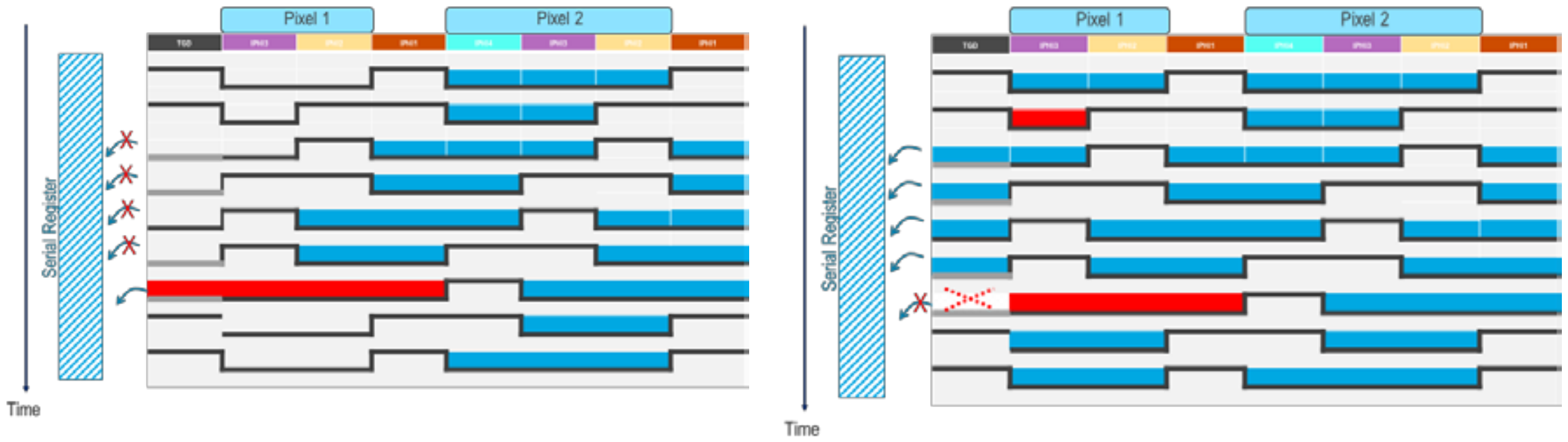

Pixel 1 is read Pixel 2 is read

**Fig. 7.** Charge transfer on the Teledyne-e2v CCD-231 using the recommended voltage and clocking and assuming a node is missing.

In **Fig. 7**, there are two different paths leading to the serial register during the line transfer cycle. If one is followed for lower intensities, and the other for higher intensities, and both of them for intensities that fall in between, this mechanism would partially explain the dip in the PTC.

## 2.5 Structural traps

An explanation for why charges would be flowing or not at different signal intensities is still needed. It requires a phenomenon involving a threshold.

One possibility is that a inhomogeneity in the electric field at the edge of the image area generates a barrier preventing charges to reach the serial register. It is supposed that this barrier collapses as soon as the first charges fall down in the serial register.

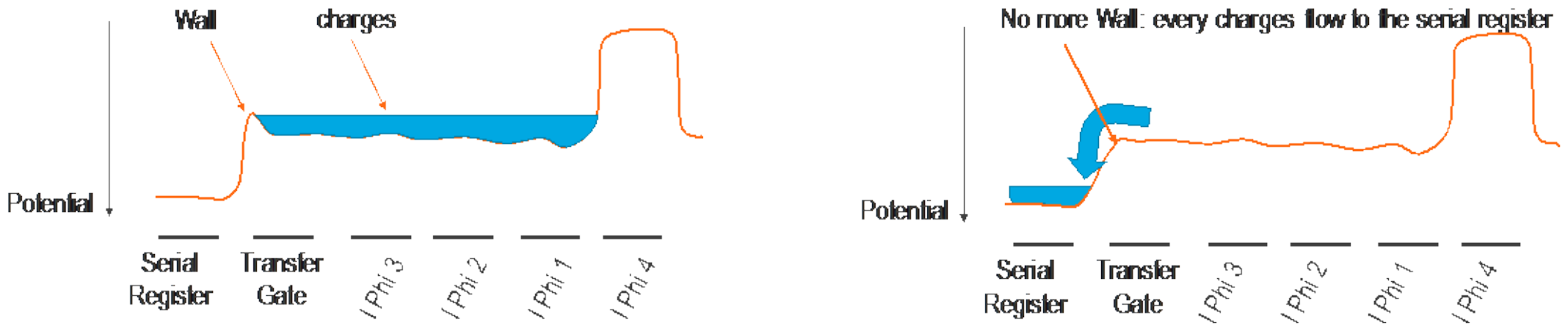

**Fig. 8.** Hypothetic field shaping leading to a level-sensitive mechanism.

## 2.6 New waveform and result

If an electrode is missing it is still possible to transfer charges correctly by modifying the clocking scheme as follows. Basically, rise and fall of the transfer gate are simply shifted to occur earlier. This way, charges can flow into the serial register only once.

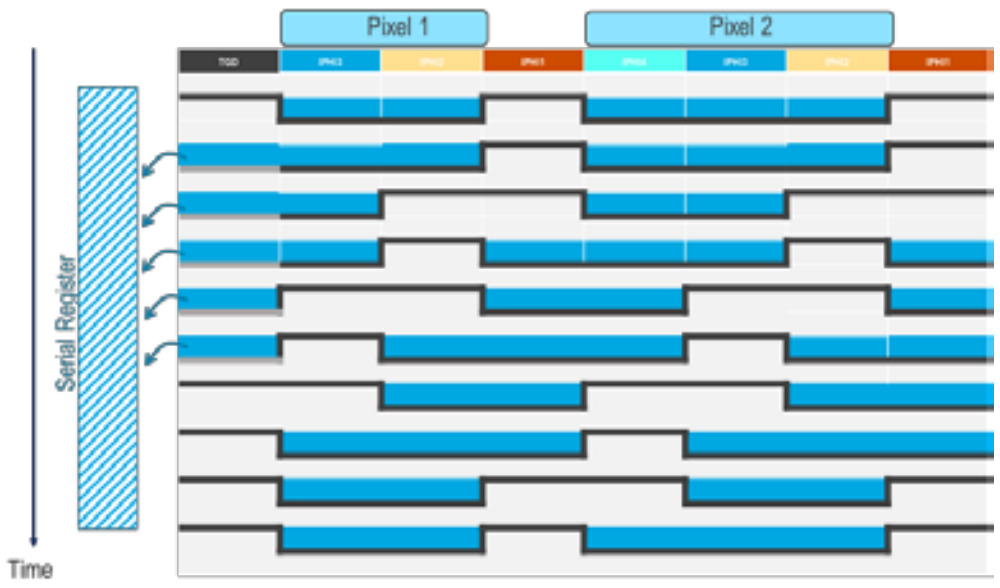

**Fig. 9.** New waveform proposed to fix the Downing Dip assuming that the last node is missing.

The resulting PTC and analysis of the pixel drifting are shown in **Fig. 10**. The downing dip and the image shift are both eliminated!

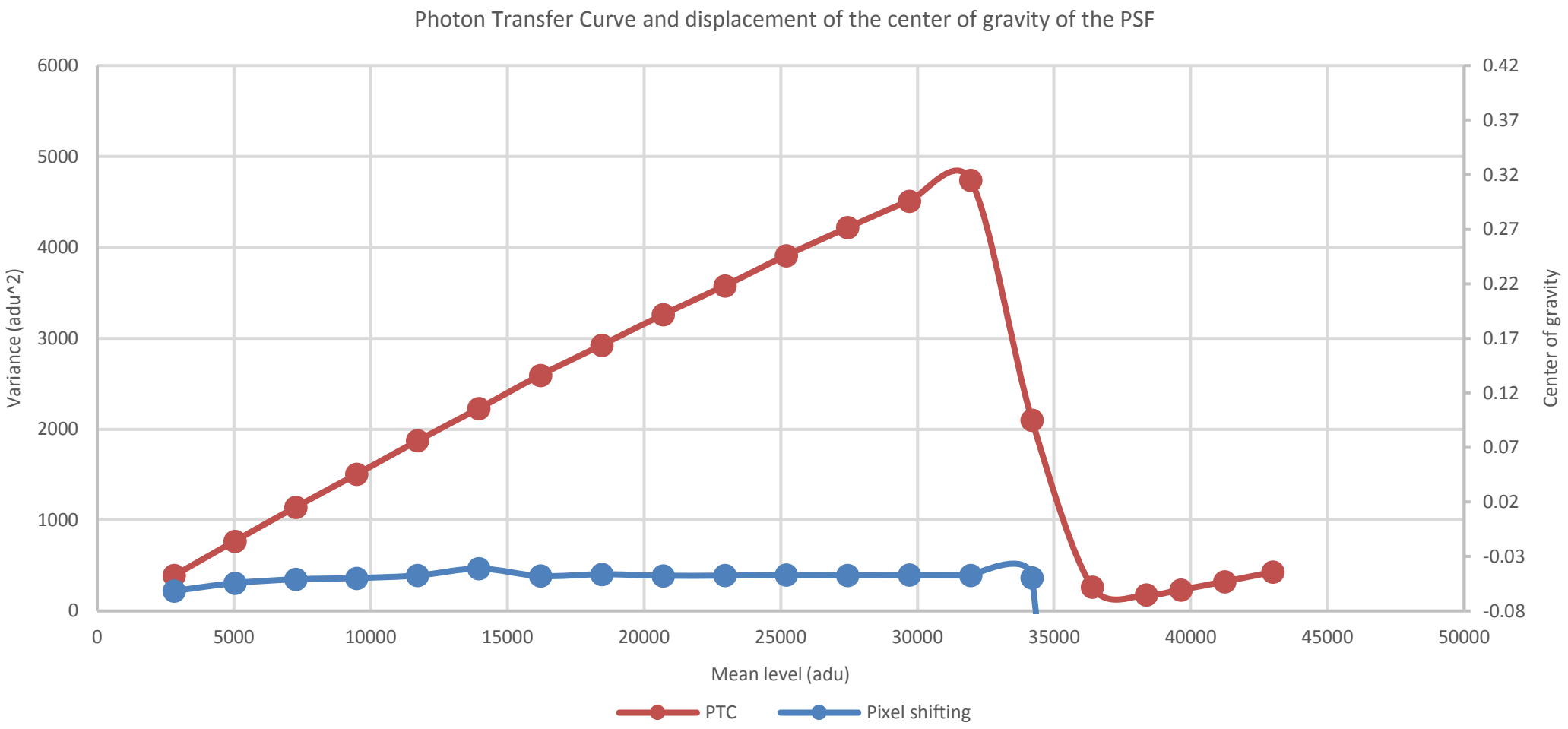

**Fig. 10.** No Downing Dip in Photon Transfer Curve, no pixel shifting using the new waveform.

In the “concurrent clocking” scheme presented by Pavan Bilgi [6] charge is transferred from last line to the serial register by setting serial clock high level more positive that Transfer Gate high which in turn is higher the parallel clock high level. Electrons then flow to the serial register under the influence of the ascending series of voltages as soon as transfer gate is lowered, without need to change the state of the parallel clocks. While the motivation was to reduce the line transfer time to just a few tens of microseconds in which TG is pulsed high, this clocking scheme has the effect of eliminating Downing Dip since the parallel clocks remain stationary and thus a missing electrode has no effect.

## 2.7 Discussion

As shown in **Fig. 11**, the electrode map of the CCD231-C6 is not symmetrical. There are 3 storage phases between the Transfer Gate and the Barrier phase on the side of the upper amplifier whereas there are only two storage phases between the Transfer Gate and the barrier phases on the lower side.

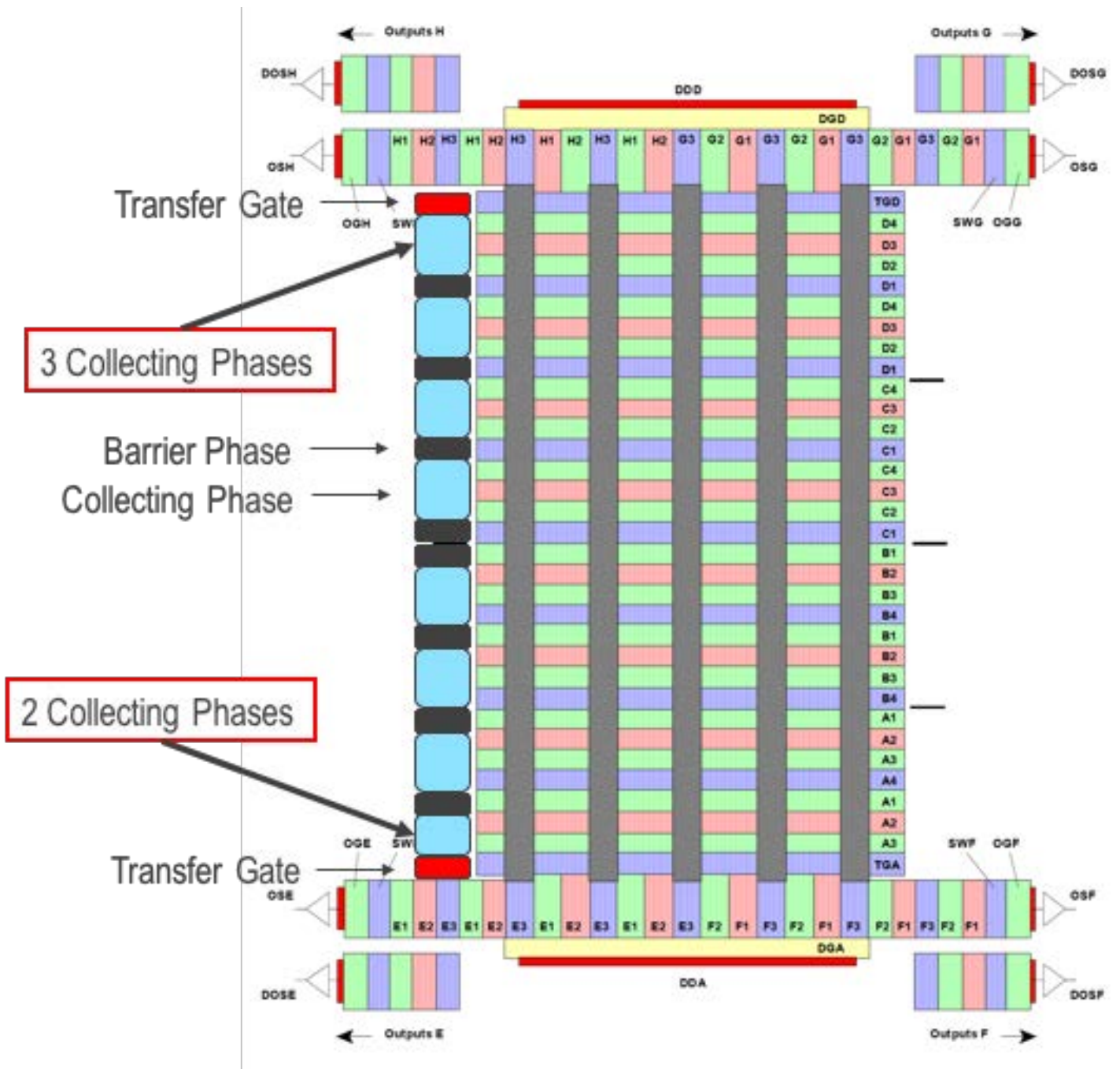

**Fig. 11.**Node mapping of the CCD-231.

The pinout of the die and the two connectors are perfectly symmetrical. Reversing the die in the package will not be noticeable when using the recommended voltages, because the Downing Dip appears only when using the sensor with lower than standard parallel clock swing.

Thus is seems plausible for the die to have been installed reversed by 180 degrees with respect to the package.

## 2.8 Another charge transfer problem

Careful analysis of the PTC generated with the new clocking highlights a very subtle dip around 66,000 e. The peak in the auto-correlation is broadened indicating smearing of charge into adjacent pixels, only in the vertical direction, and only where the dip occurs.

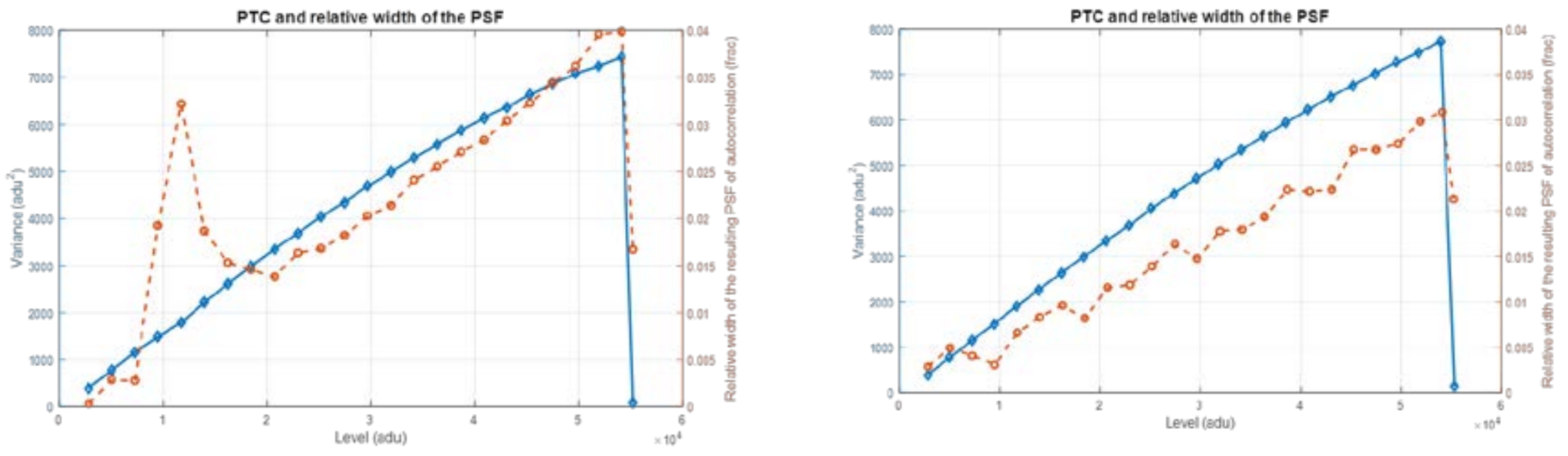

Subtle Dip and spreading of the PSF

No dip, no PSF's spreading with new voltage

**Fig. 12.**PTC and relative width of the PSF before and after applying new voltage. The PSF is increasing with higher light level because of the Brighter-Fatter effect.

Because all lines are affected equally, we hypothesized that a structural trap existed at the interface between the Transfer Gate and the serial register. In support of this we find that increasing the Serial Clock High voltage by 1V compared to the Transfer Gate is enough to fix this problem. (**Fig. 12**).

## 3 Surface Full Well

Pictures taken during a first engineering run of the WaSP camera at Palomar Mountain show long tails below and across some, *but not all* bloomed stars.

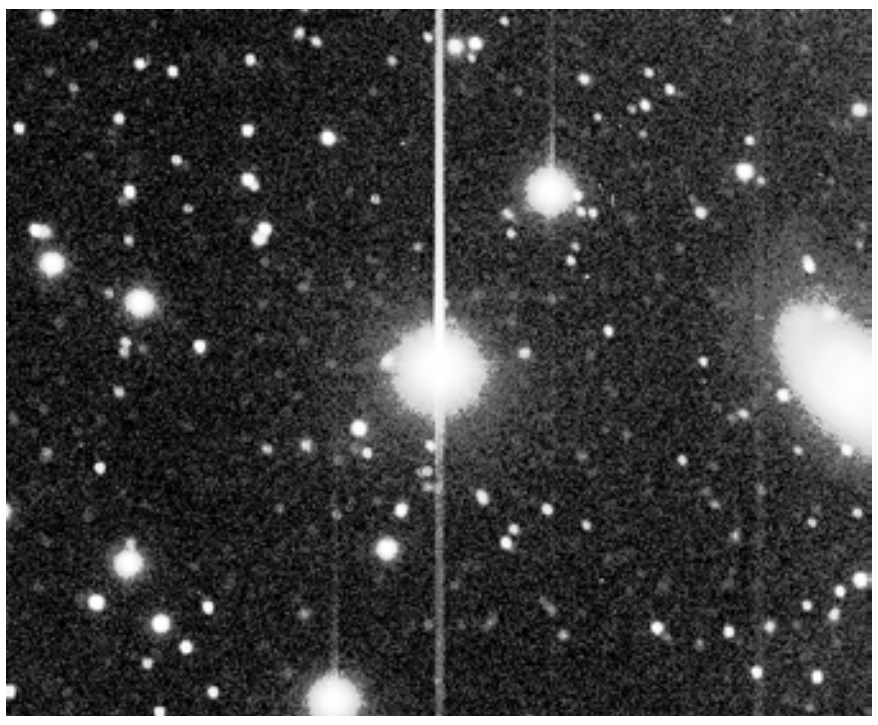

**Fig. 13.** Bloomed star on the Field of View showing vertical tails.

These tails are believed to be caused by charges trapped on the Si/SiO2 surface during the exposure. A slow release of trapped charge during the readout creates the tail following the source, while the tail seen preceding it is due to the release of charge during the previous frame readout.

Such a tail is highly detrimental to image quality since a single over-illuminated pixel affects not only its neighborhood (as for by bloomed charges) but the entire column within a quadrant.

### 3.1 Surface Full Well versus Blooming Full Well

To study this phenomenon, a pinhole was located about 300 pixels from the center of the sensor in such a way it is possible to measure the level of the tail on both image area and parallel overscan. The level of the tail is referenced to an unaffected area of the overscan as shown in **Fig. 14**.

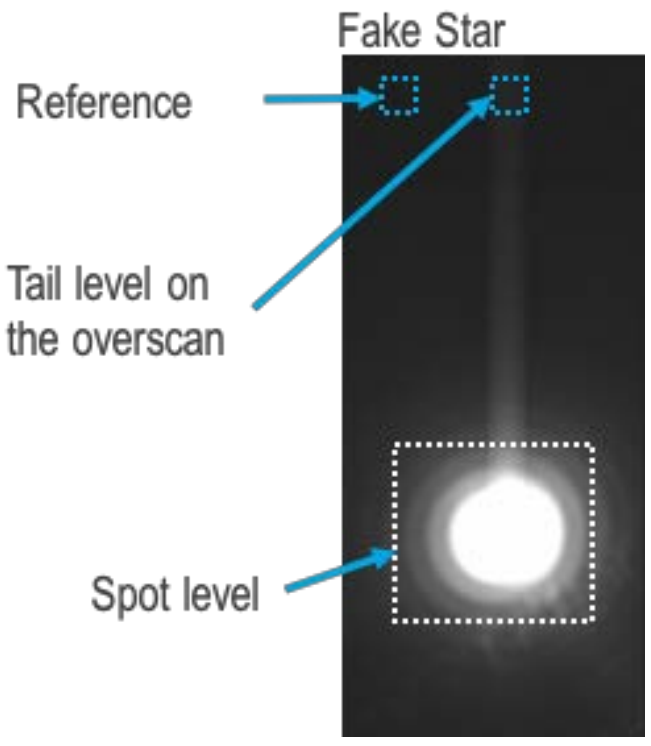

**Fig. 14.** Pinhole setup with spot and visible tail.

This setup was used to plot the levels of the pinhole and the tail using the initial 0 to 10 volts clocking. Increasing the integration time allows us to reach over-illumination level not previously probed when plotting Photon Transfer Curves.

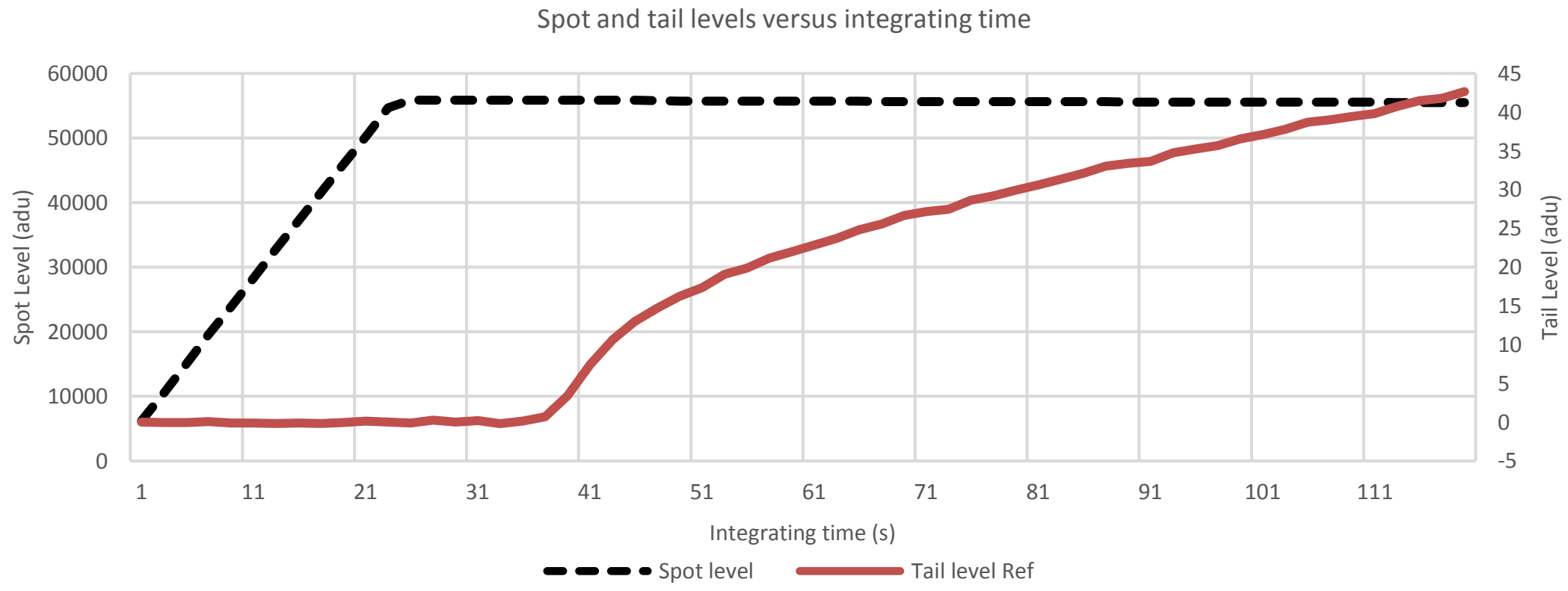

**Fig. 15.** Pinhole setup with spot and visible tail.

**Fig. 15** shows that tails appear for levels roughly twice the blooming level.

Surface Full Well was not expected after Blooming! The 1-D representation of the electric field from Janesick [2] forbids such behavior since the charges are expected to spill over the barrier phase and reach neighboring pixels before reaching the surface.

Using a more complex 2-D model simulation tool and a pixel model of the CCD250, D.P. Weatherill et al. [7] show that Surface Full Well can indeed be reached even after Blooming Full Well.

## 3.2 Inverted clocking

A classic way to eliminate the charge tail after pixel reaching Surface Full Well is to use inverted clocking. As each phase is inverted, holes drawn from the channel stops recombine with charge trapped at the surface before it is collected in the well.

We found that it was necessary to invert to eliminate the charge tails.

Since flatness of the PTC decreases with a lower common mode voltage on the image area. Minus 5 volts to 5 volts is a good trade-off.

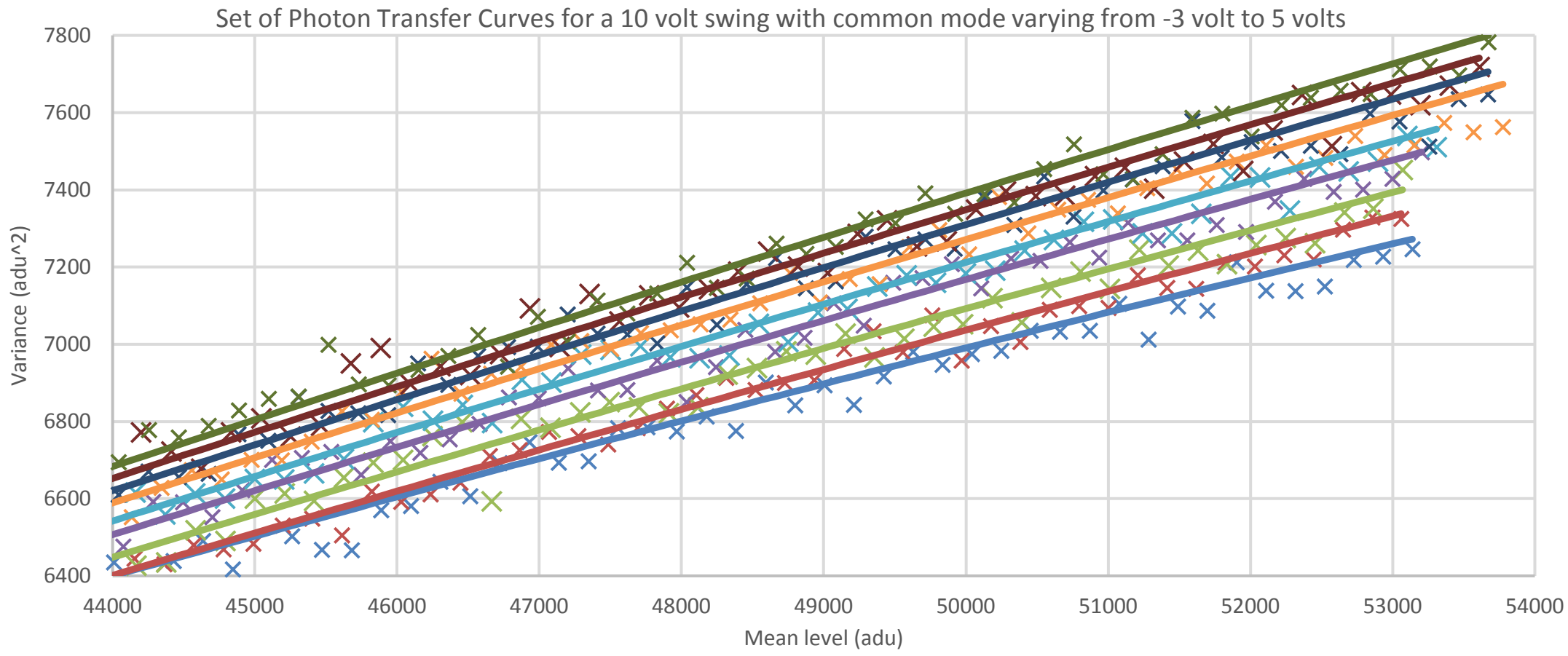

**Fig. 16.** Increase of the steepness of the PTC with higher common mode voltage (zoom on the end of the PTC). From a voltage swing of -8V to 2V in blue to 0V to 10V in dark green. (red:-7V to 3V, light green:-6V to 4V, purple: -5V to 5V, light blue: -4V to 6V, orange: -3V to 7V, dark blue: -2V to 8V, brown: -1V to 9V).

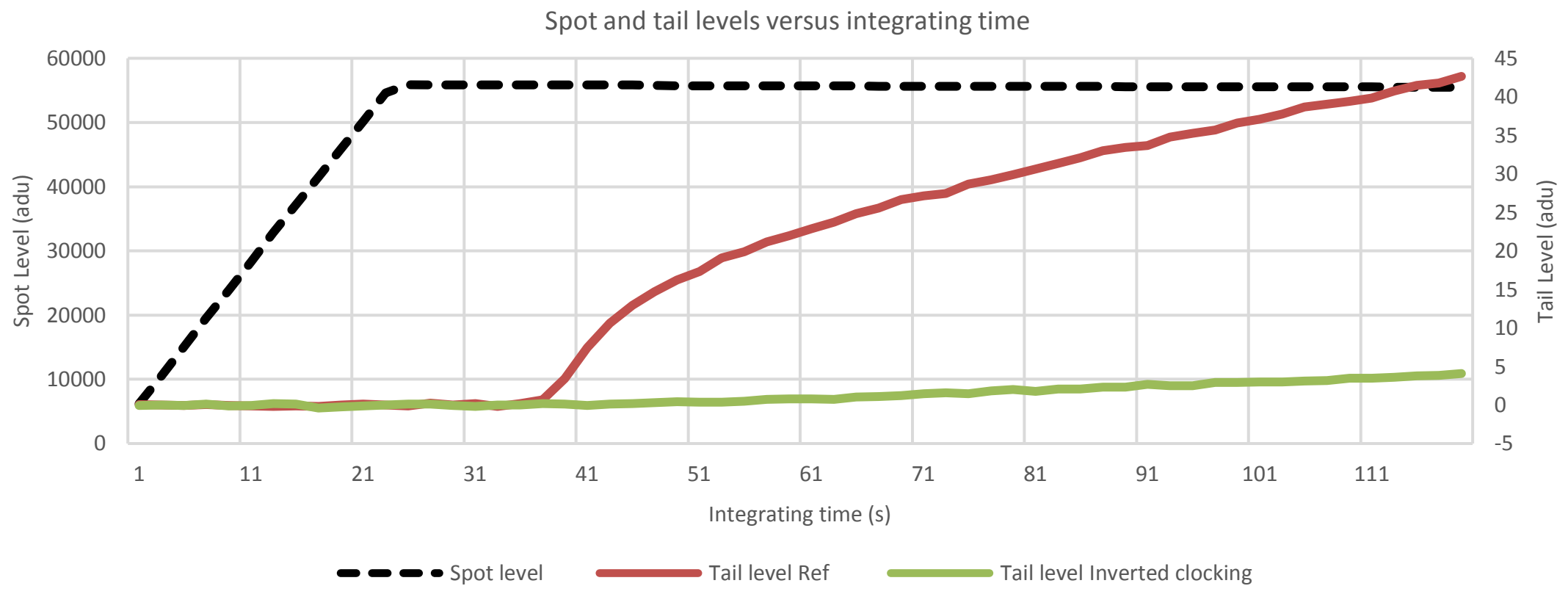

**Fig. 17.** Pinhole setup with spot and visible tail. Inverted clocking (-5V to 5V) reduces intensity of the tail.

As shown in **Fig. 17**, the Surface Full Well is highly reduced using even just partially inverted clocking.

## 3.3 Bi-clocking

The main drawback of inverting clocks during exposure is an increase of the brighter-fatter effect. The obvious solution is bi-clocking scheme in which parallel phases are held at high common mode voltage during exposure to limit the brighter-fatter and reduce lateral charge diffusion by extending the electric fields deeper into the silicon, then during readout a lower common mode voltage is sued to both keep charge away from the Si-SiO2 surface during charge transport and to clean up charge trapped there during the exposure. This method shows excellent results since the tail is reduced by a factor of about 20 for the longer exposure time.

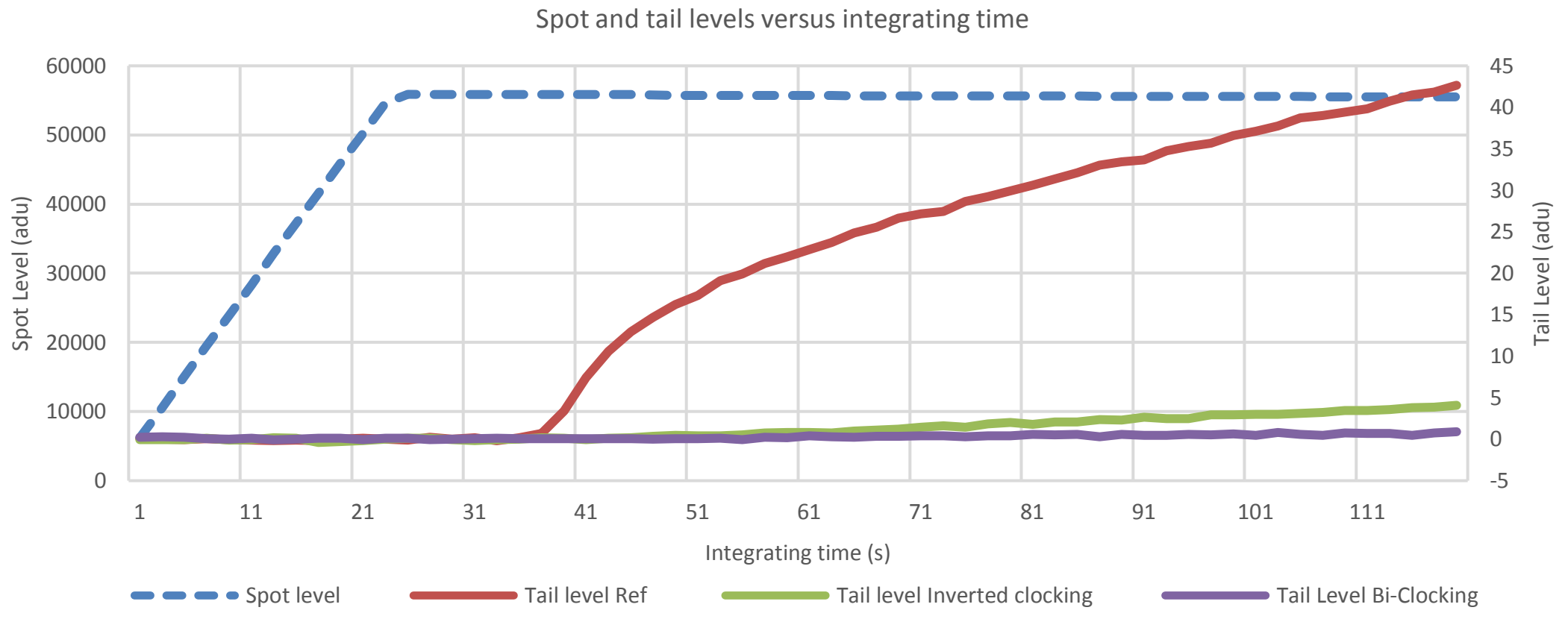

**Fig. 18.** Levels of spot and tails. In purple: using the bi-clocking with 3V to 13V during exposure and -8V to 2V during readout.

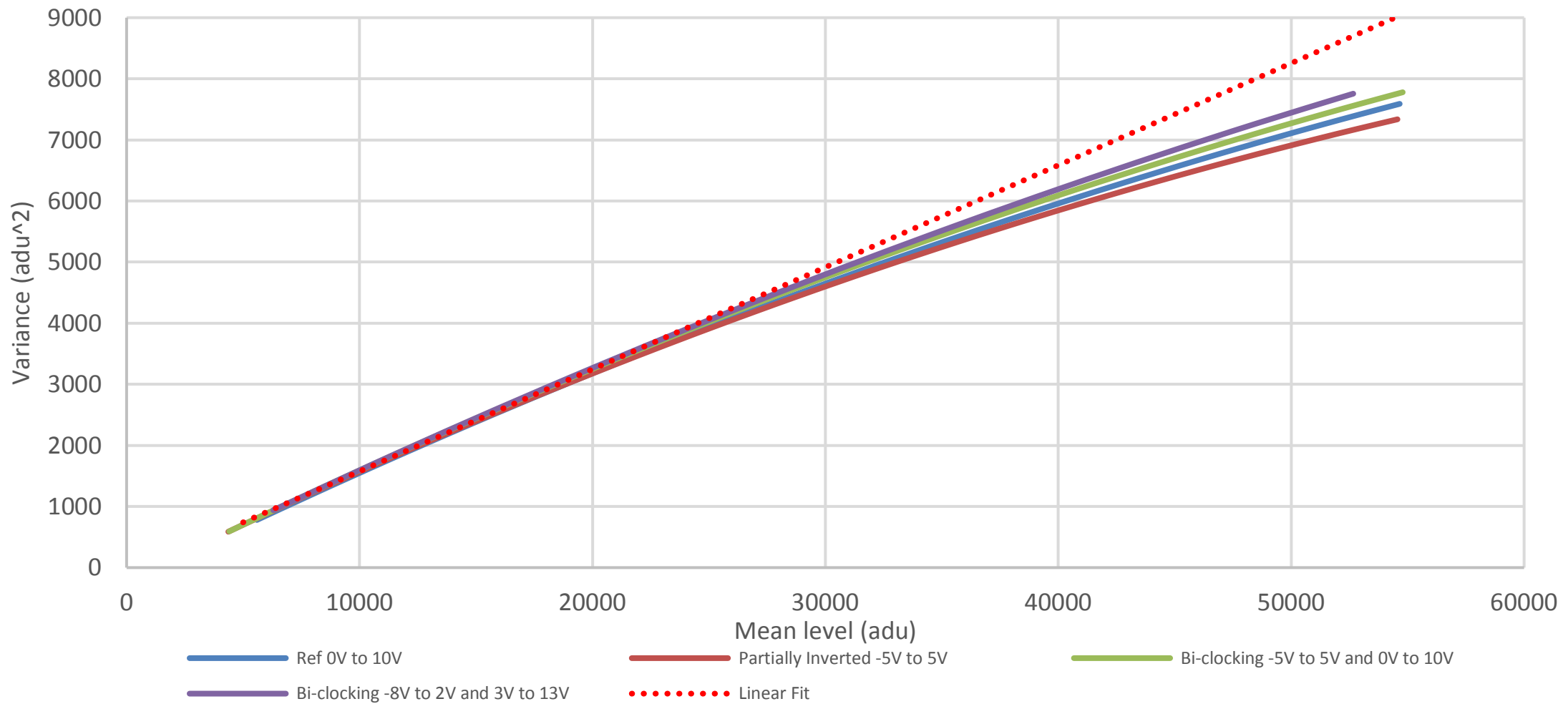

**Fig. 19.** Photon Transfer Curves using classical 0-10V clocking during both exposure and readout (blue), weakly inverted clocking -5V to +5V (brown), and bi-clocking with -8V to2V swing during readout and 0-10V during exposure (green) or 3V to 13V during exposure (purple).

The fact that PTC is even steeper when using the bi-clocking was unexpected.

The best PTC straightness is reached holding clocks at 3V for one barrier phase and 13V on the three storage phases during the exposure, then using fully inverted clocking, swinging -8V to 2V, during charge transfer. Currently, we are limited by the driving capabilities of our controller but a higher voltage on the image area during the exposure could be tried and could eventually lead to an even flatter PTC.

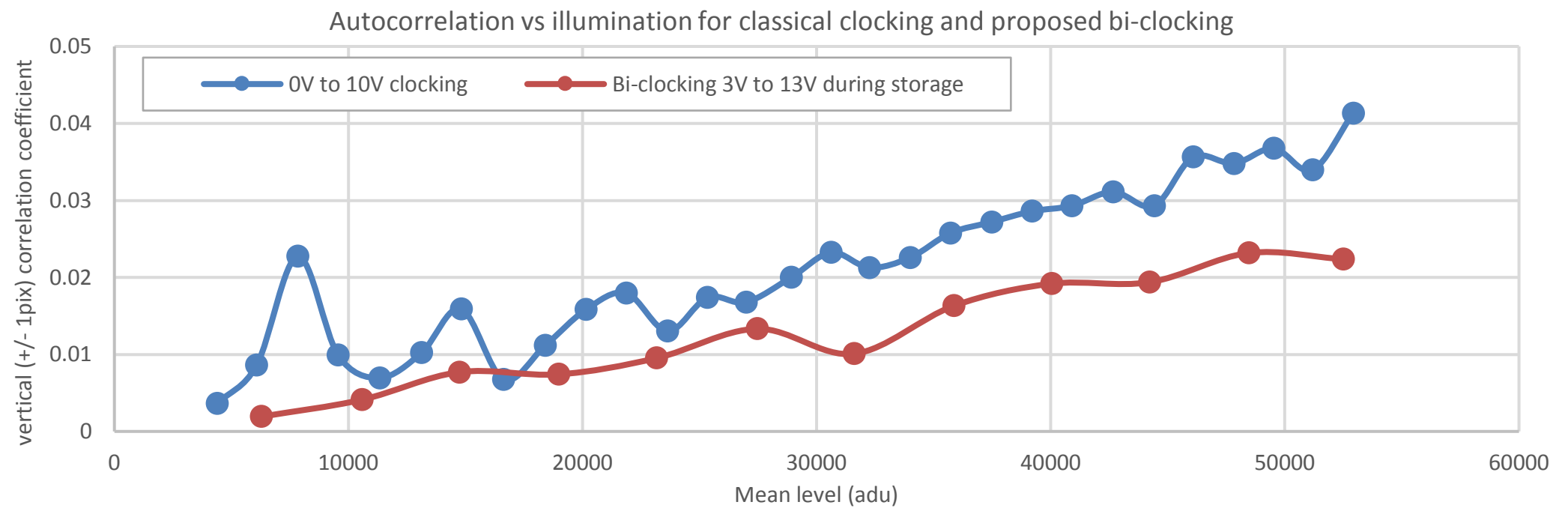

**Fig. 20.** Correlation coefficients of flat-field pixels pair as a function of illumination of the flat-field, using the classical 0V to 10V clocking or the Bi-clocking.

The reduction in correlation pixels appears to be due to the relative strength imposed electric fields and those fields generated by stored charge. The fact that correlation scales almost linearly with signal supports the charge repulsion explanation (brighter-fatter effect) as opposed to surface trapping which will show an onset threshold on the PTC.

## 4 Pixel Response Non-Uniformity

Pixel Response Non-Uniformity measurements have been made following the method described by Smith and Rahmer [8]. Fifteen flat fields are co-added, each with approximately 250k electrons per pixel, so that shot noise per pixel is reduced to ~0.05% of the mean, compared to a 0.3 % rms PRNU. The high frequency PRNU structure is extracted by dividing the flat field by a 9x9 boxcar-filtered version of itself.

This size boxcar is chosen to remove both the illumination gradient in the flats and mid frequency ripples induced by the laser annealing of the boron implant, which are not relevant to this study.

### 4.1 2 phases vs 3 phases

A map of the pixel area was generated using the recommended 2 collecting phases clocking. Vertical stripes are visible along the column at the center of the chip. The vertical stripes are still evident at the center of the chip using a split serial transfer or a single serial transfer, which indicates they are dependent on position and not readout direction or timing. Using 3 phases for integration suppressed these vertical stripes. The phenomenon leading to this is not understood.

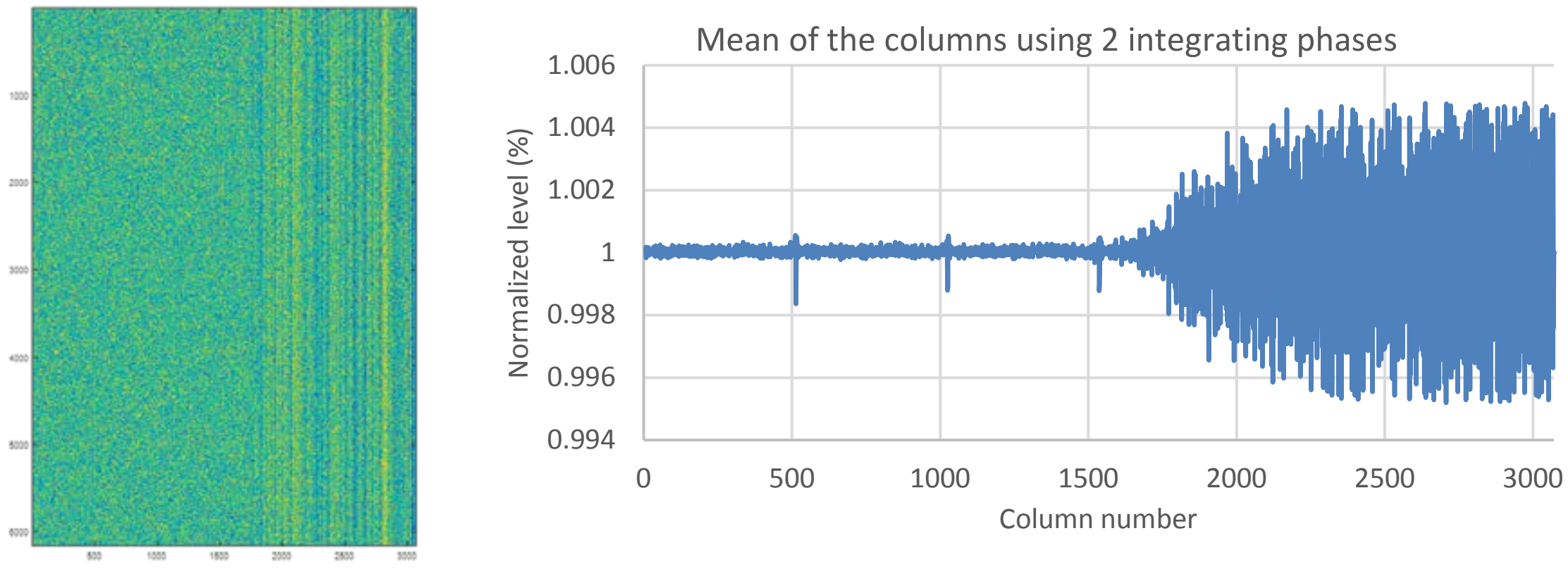

**Fig. 21.** Pixel area map (Left side) using a 2 phases storage clocking. Columns along the center of the chip show excess of noise. Right panel shows the average level of each column.

Even when the standard deviation of the PRNU map is computed over a small sub-window chosen to avoid the vertical stripes, integrating under 3 phases is found to be beneficial reducing PRNU from 0.38% to 0.28% in areas with no dust or cosmetic defects.

**Table 1.** PRNU at red wavelengths for integrations under 2 phases or 3 phases.

| | 2 storage phases | 3 storage phases |
|---|---|---|
| PRNU in red wavelength | 0.38% | 0.28% |

### 4.2 Lithographic mask

Averaging the pixel area map along the horizontal and vertical directions highlights the lithographic mask stitching in the image area. Teledyne-e2v is using a lithographic mask of 256 by 512 pixels. Since the sensor’s geometry is 6144x6160 the 8 lines closest to the serial register are printed with a different mask. The line widths at the stitching boundaries are accurate to about ±0.1% which is equivalent to 15 nm.

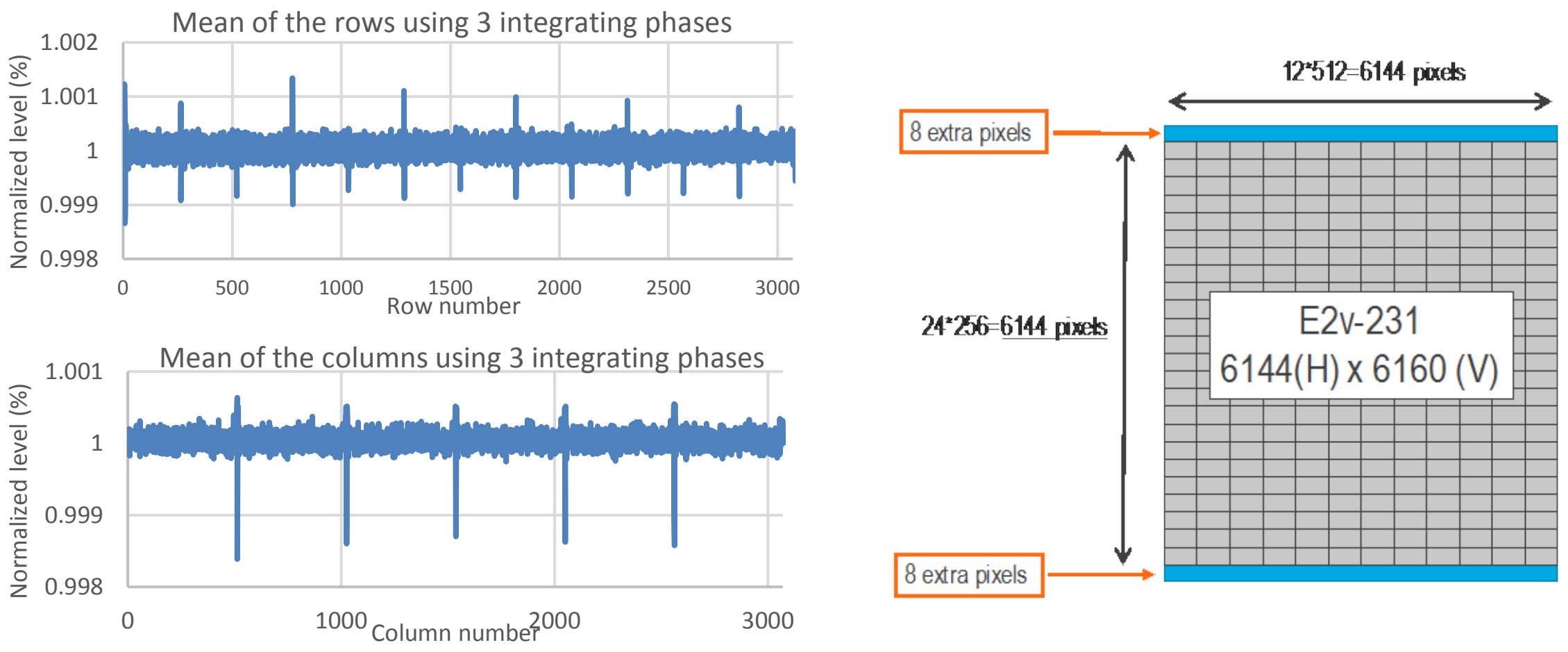

**Fig. 22.** Lithographic mask stitching revealed when averaging the PRNU map on the horizontal and vertical directions.

## 4.3 Wavelength dependence

A monochromatic illuminator made with selectable LEDs has been used to assess the wavelength dependency of the flat field structure. High intensity flats are coadded to reduce shot noise to be well below the patterns observed.

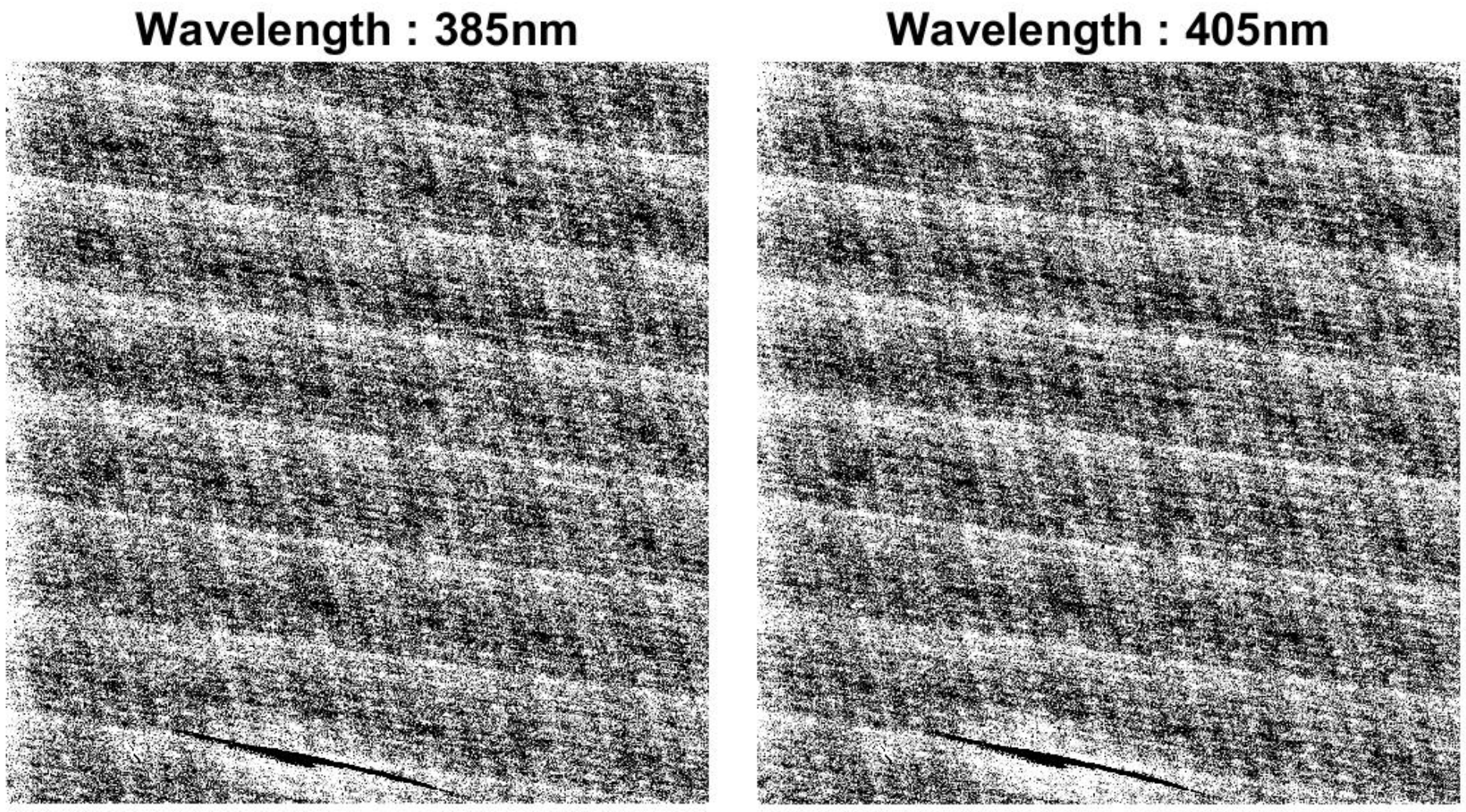

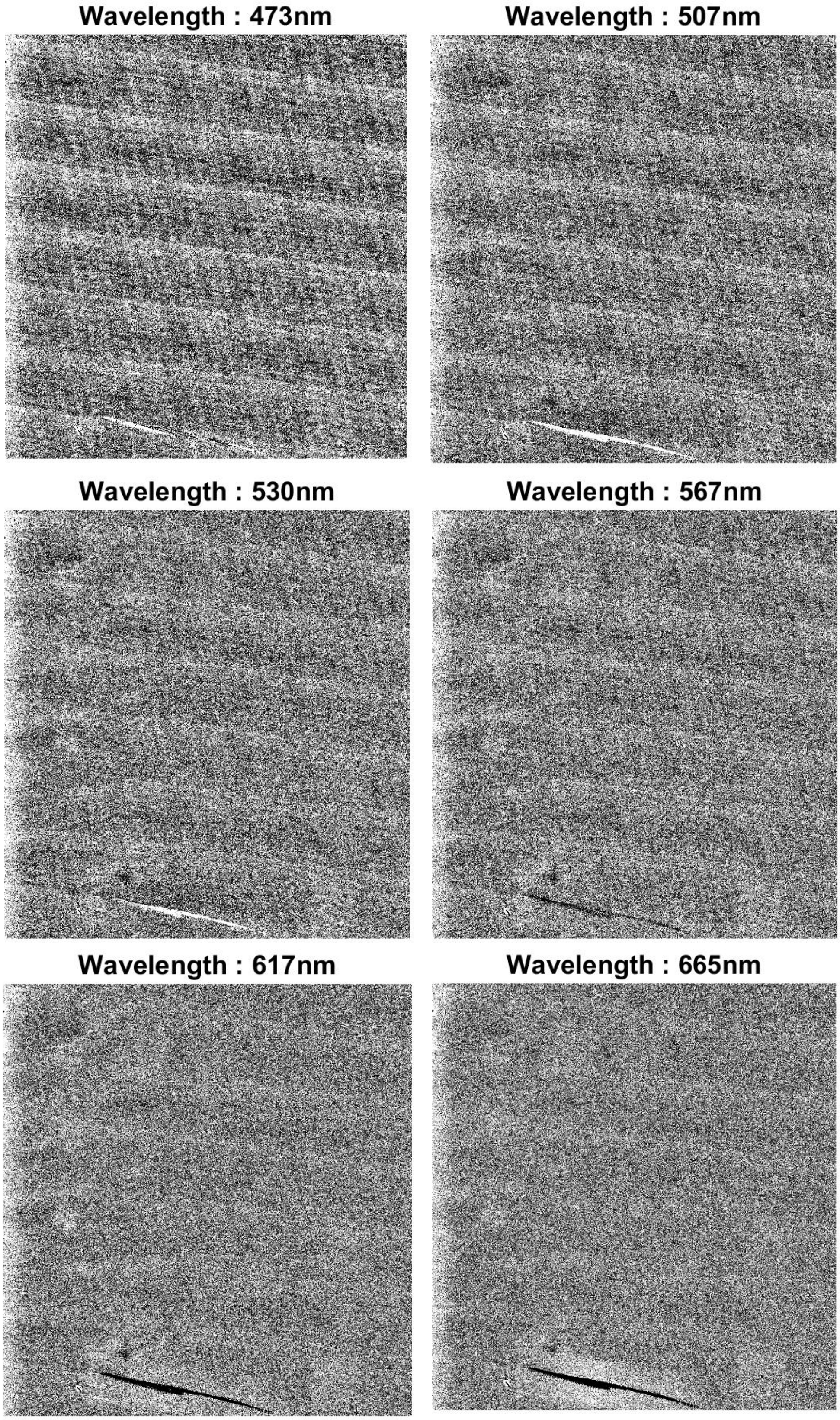

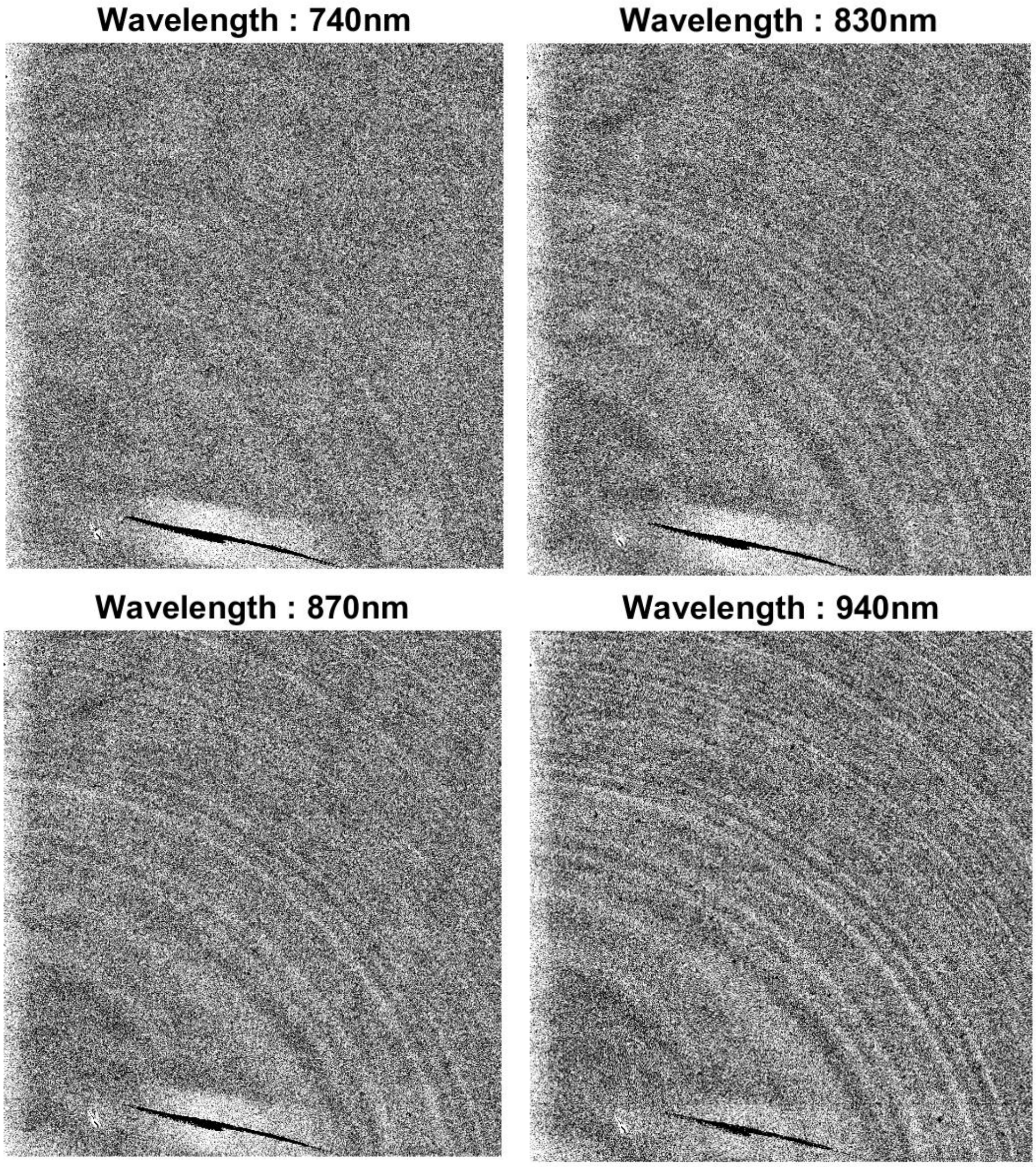

**Fig. 23.** Patterns on the upper right quadrant of the CCD-231 for monochromatic illumination from 385nm to 940nm.Black level is 0.5% below mean level and white level is above 0.5% of the mean level

Diagonal crosshatch pattern is due to the laser annealing of the Boron implant at the back surface and thus has strongest effect at shortest wavelengths. The horizontal "scratch" changes from dark to bright and back implying that it is an anti-reflection coating defect. The rings appearing at longer wavelengths are always centered on the wafer and are believed to be due to lateral electric fields produced by impurity atoms acting as charge donors or acceptors in the depleted thickness of the CCD. These lateral electric fields modulate pixel boundary positions and thus pixel area, to produce these very weak structures in these flat fields. A similar effect is seen very prominently in thick fully depleted CCDs. In this instance, it is puzzling that the rings are not evident at shorter wavelength since all photogenerated charge passes through the depletion region where the wells reside. The presence only at the longest wavelengths suggest that the effect is occurring between the wells and the front surface.

In **Fig. 23**, a pattern on scales larger much than 9 pixel have been suppressed using the moving boxcar filter.  This was done since the illumination was not flat enough to display the low contrast pattern observed.

In **Fig. 24**, flats from ZTF with much better illumination uniformity reveal a wafer scale gradient where the a central zone, about 4000 pixels across has greater red response. This corresponds at least approximately to the metrology for surface height and thus may reflect enhanced long wavelength QE due to greater thickness.

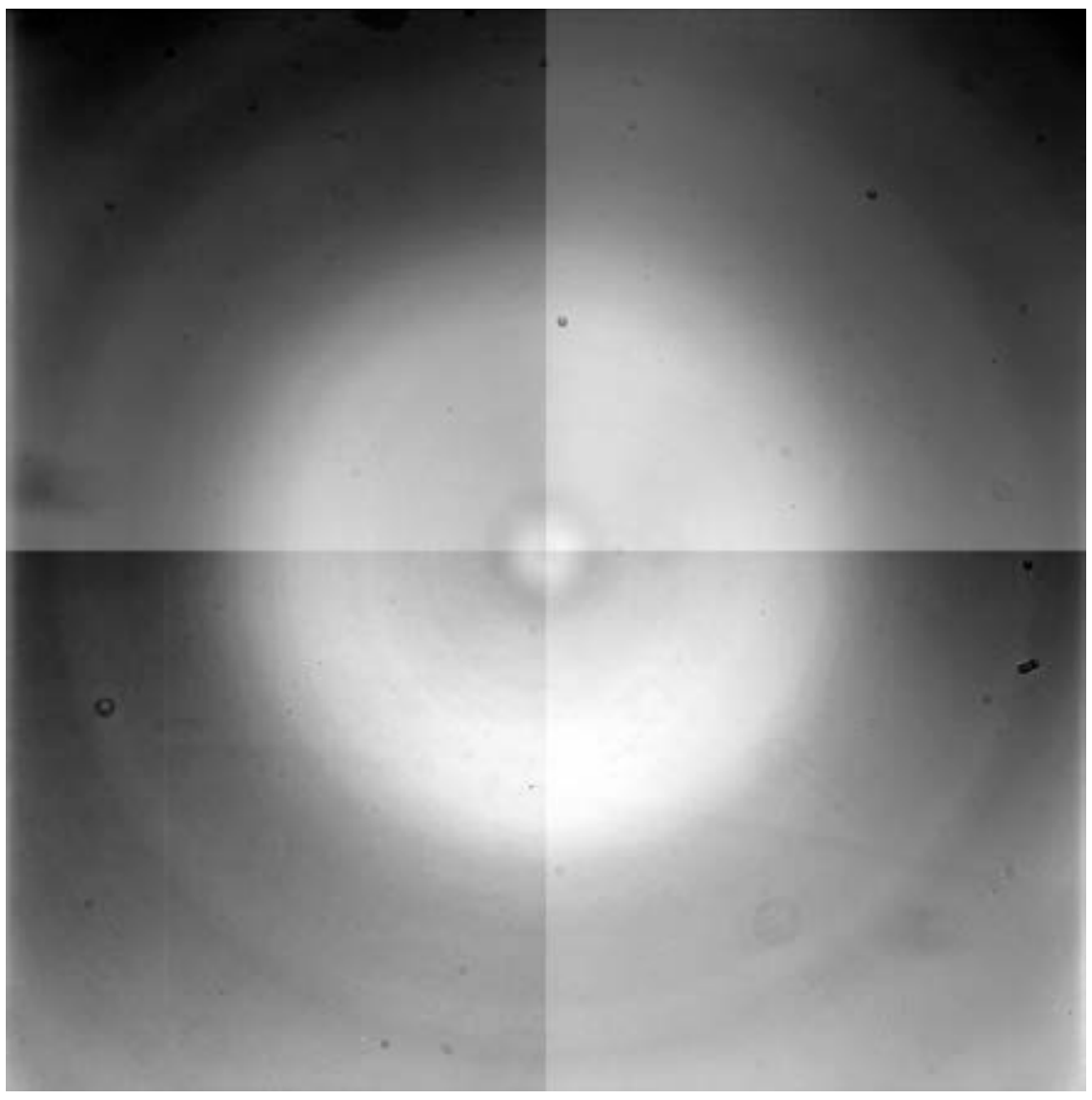

**Fig. 24.** Flat field on ZTF. The better illumination uniformity reveals low-frequency pattern.

## 5 Conclusion

While attempting to make the Photon Transfer Curve of our CCDs straighter, we discovered and understood several charge transport issues for the Teledyne-e2v CCD231-C6.

The Downing Dip seen in the PTC at about 75% of the blooming threshold was eliminated by using clocking that assumes the die is inverted with respect to the datasheet.

A subtler noise correlation at around 70 ke- was eliminated by increasing the difference in high voltage between the transfer gate and the serial register.

Positively biased clocks produce straighter PTC indicating a weaker brighter-fatter effect and leads to the expectation that lateral diffusion will be reduced.

Tails were found behind stars several times brighter than the blooming threshold, leading us to the explanation that Surface Full well can in fact occur beyond Blooming threshold.  By adopting very positively biased clocks during the exposure, which tend to enhance surface trapping, and conventional inverted clocks during readout we both cured the charge trails from surface traps and minimized the lateral charge diffusion and associated shot noise correlation that is symptomatic of the brighter fatter effect.

Integrating using 3 storage phases improved not only the PRNU over a clean area on the sensor but drastically improved photometry over half of the detector by suppressing vertical stripes seen in PRNU maps.

The clock sequences and voltages proposed are improving the image quality of the CCD231-C6 sensors in both WaSP and ZTF cameras.

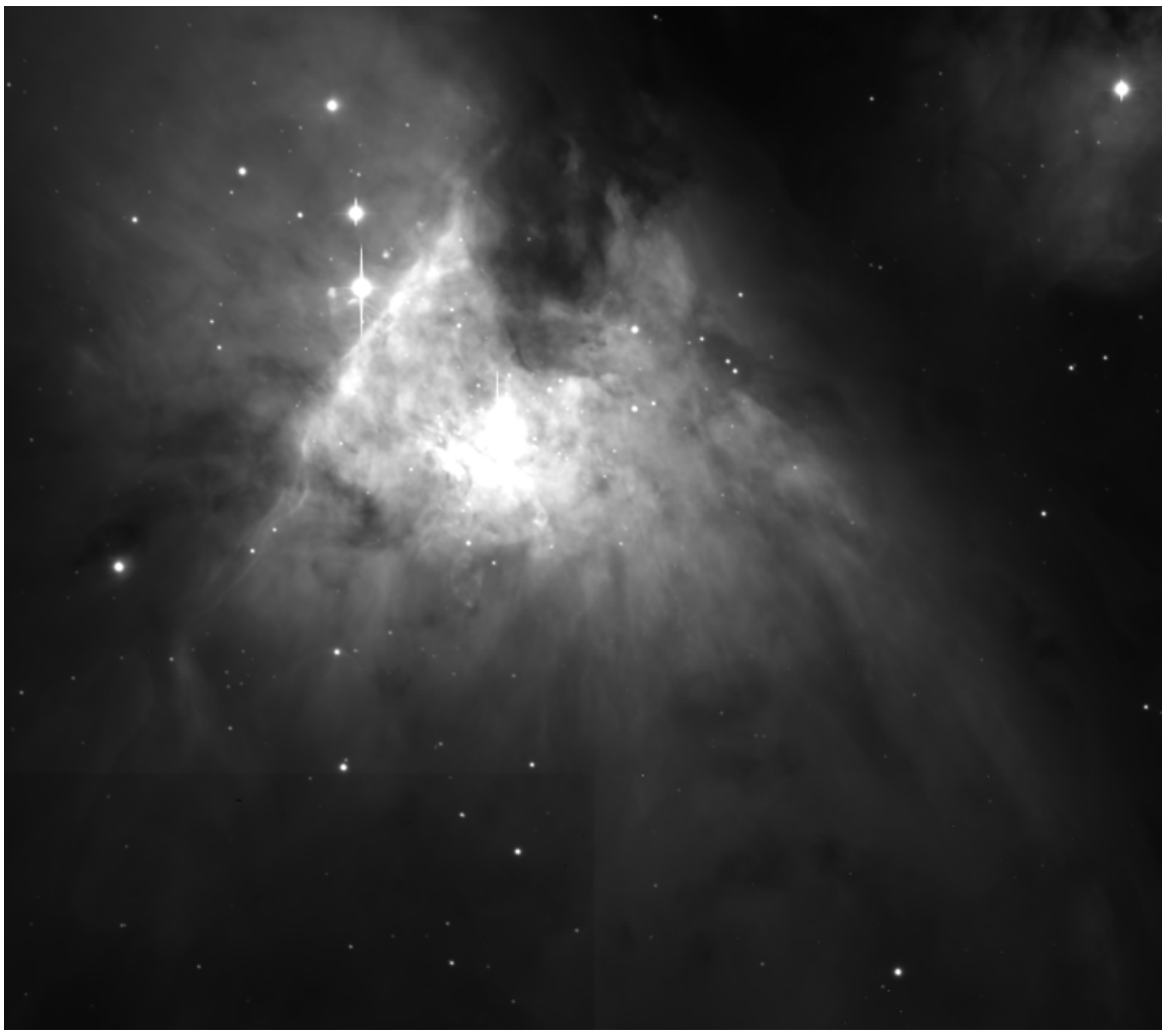

**Fig. 25.** M42 Nebula taken at the Palomar Observatory during an engineering run on November 2017. Bi-clocking is implemented and no tails are seen behind over illuminated stars.